\documentstyle[preprint,tighten,pra,aps]{revtex}
\begin{document}
\tightenlines
\draft

\preprint{UTF 414}

\title{Topological Black Holes in Weyl Conformal Gravity}

\author{Dietmar Klemm\footnote{email: klemm@science.unitn.it}}
\address{Dipartimento di Fisica, Universit\`a  di Trento, Italia}

\maketitle\maketitle\begin{abstract}
We present a class of exact solutions of Weyl conformal gravity, which
have an interpretation as topological black holes. Solutions
with negative, zero or positive scalar curvature at infinity are found,
the former generalizing the
well-known topological black holes in anti-de Sitter gravity.
The rather delicate question of thermodynamic properties of such
objects in Weyl conformal gravity is discussed; suggesting that
the thermodynamics of the found solutions should be treated within the
framework of gravity as an induced phenomenon, in the spirit of
Sakharov's work.
\end{abstract}


\maketitle

\section{Introduction}

Topological black holes \cite{amin,mann,lemos,zhang,huang,vanzo}
represent an interesting domain of
general relativity. They are expected to occur naturally in
any future theory of quantum gravity involving topological transition
processes, and they may provide further ground to test string theory ideas in
black hole physics \cite{kv}. A necessary condition for their existence
seems to be the presence of a negative cosmological constant, i.~e.~they occur
in anti-de Sitter (AdS) gravity. Indeed, Hawking's theorem \cite{hawk1}
states that under the assumption of asymptotic
flatness and positivity of matter-energy, the event horizon of a
black hole is always spherical.\\
Now in the action of Weyl conformal gravity there is no room for a cosmological
constant, as this would introduce a length scale and hence break
conformal invariance. This raises the question if objects like
topological black holes can nevertheless exist in such a theory of
gravitation. (Note in this context that recently there have been found
exact solutions of other higher derivative theories, which represent
black holes in an AdS background \cite{bonanno}. In this work, the
cosmological constant $\Lambda$ arises as a product of the integration
procedure, and is not taken as an ingredient of the action. However,
the field equations are solved only for special values of $\Lambda$).\\
In this letter we want to investigate if Weyl conformal gravity
admits solutions representing topological black holes. We do not
discuss the advantages or drawbacks
of this alternative theory of
gravitation, as there is a vast list of literature dedicated to this subject.
In any case, it seems interesting to consider black holes in Weyl
gravity, because the requirement of conformal invariance at the
classical level leads to a renormalizable gauge theory of gravity,
which permits a consistent picture of black hole evaporation \cite{hassl}.\\
The remainder of this paper is organized as follows:\\
In section \ref{action} we review the gravitational action, the
field equations and the most general spherically symmetric solution
of Weyl conformal gravity.\\
In section \ref{sol} exact solutions having an interpretation
as black holes with non-trivial topology are derived. As a special case we will
recover the well-known topological black holes in AdS gravity, but we
also find other cases which, in contrast to the former,
have non-negative scalar curvature at infinity.\\
In section \ref{thermo} we discuss the black holes' thermodynamic
properties, which will result to be
a rather ticklish concern,
e.~g.~applying Wald's definition of thermodynamic variables for
an arbitrary, diffeomorphism-invariant Lagrangian \cite{wald},
seems to yield only
trivial results in the case of black holes in Weyl conformal gravity.
We therefore suggest a possible treatment of the thermodynamics of the
found solutions by the idea of gravity as an induced phenomenon
\cite{sakh}.\\
Finally, our results are summarized and discussed in section \ref{disc}.

\section{Weyl Conformal Gravity} \label{action}

The conformally invariant action under consideration is
\begin{eqnarray}
I_W &=& \alpha\int d^4x \sqrt{-g} C_{\mu\nu\rho\lambda}C^{\mu\nu
      \rho\lambda} \nonumber \\
   &=& 2\alpha\int d^4x \sqrt{-g} [R_{\mu\nu}R^{\mu\nu} - \frac{1}{3}
       (R^{\mu}_{\; \mu})^2], \label{act}
\end{eqnarray}
where $C_{\mu\nu\rho\lambda}$ is the Weyl tensor and $\alpha$
is a dimensionless parameter, which is usually chosen to be positive,
in order to have a positive definite euclidean action and an acceptable
newtonian limit.\\
The vacuum field equations following from (\ref{act}) read
\cite{dewitt,mannhkaz,mannh}
\begin{equation}
W_{\mu\nu} = W_{\mu\nu}^{(2)} - \frac{1}{3}W_{\mu\nu}^{(1)} = 0,
\label{fieldeq}
\end{equation}
where $W_{\mu\nu}^{(1)}$ and $W_{\mu\nu}^{(2)}$ are given by
\begin{eqnarray}
W_{\mu\nu}^{(1)} &=& 2g_{\mu\nu}\nabla_{\beta}\nabla^{\beta}
                     R^{\alpha}_{\; \alpha} -
                     2\nabla_{\nu}\nabla_{\mu}R^{\alpha}_{\; \alpha}
                     - 2R^{\alpha}_{\; \alpha} R_{\mu\nu} +
                     \frac{1}{2}g_{\mu\nu}(R^{\alpha}_{\; \alpha})^2,
                     \nonumber \\
W_{\mu\nu}^{(2)} &=& \frac{1}{2}g_{\mu\nu}\nabla_{\beta}\nabla^{\beta}
                     R^{\alpha}_{\; \alpha} +
                     \nabla_{\beta}\nabla^{\beta}R_{\mu\nu} -
                     \nabla_{\beta}\nabla_{\nu}R_{\mu}^{\; \beta} -
                     \nabla_{\beta}\nabla_{\mu}R_{\nu}^{\; \beta} -
                     2R_{\mu\beta}R_{\nu}^{\; \beta} +
                     \frac{1}{2}g_{\mu\nu}R_{\alpha\beta}R^{\alpha\beta}.
\end{eqnarray}
As $W_{\mu\nu}^{(1)}$ and $W_{\mu\nu}^{(2)}$ vanish identically for
$R_{\mu\nu} = \Lambda g_{\mu\nu}$,
this implies that every Einstein space
also satisfies the vacuum field equations
of Weyl conformal gravity. As the topological black holes discussed in
\cite{amin,mann,lemos,zhang,huang,vanzo} are Einstein spaces (if uncharged),
they also fulfill
(\ref{fieldeq}). However,
inspecting the most general static, spherically
symmetric solution of (\ref{fieldeq}) (see below) derived in
\cite{mannhkaz},
we expect the field equations of Weyl conformal
gravity to admit also a richer class of topological black hole solutions,
which will not consist of Einstein spaces alone.
As we will see in section \ref{sol}, this is indeed the case.\\
The general static, spherically symmetric line element
resolving (\ref{fieldeq})
reads \cite{mannhkaz}
\begin{equation}
ds^2 = -V(\rho)d\tau^2 + V(\rho)^{-1}d\rho^2 + \rho^2(d\Theta^2 +
\sin^2\Theta d\Phi^2),
\label{spher}
\end{equation}
with $V(\rho)$ given by
\begin{equation}
V(\rho) = 1 - \frac{\beta(2-3\beta\gamma)}{\rho} - 3\beta\gamma +
\gamma \rho -k \rho^2,
\label{lapse}
\end{equation}
$\beta$, $\gamma$ and $k$ being constants.
(Of course multiplying (\ref{spher}) by a conformal factor depending
only on $\rho$, one again obtains a static, spherically symmetric solution).
One notes that for $\gamma = 0$, (\ref{lapse})
yields the Schwarz\-schild-, Schwarz\-schild-(anti-)de Sitter-, and
(anti-)de Sitter spacetimes, depending on the choice of the parameters,
but for $\gamma \neq 0$ there arises an additional potential linear in $\rho$.

\section{Topological Black Hole Solutions}
\label{sol}
To obtain topological black holes, we analytically continue (\ref{spher})
according to
\begin{equation}
\tau = it, \qquad \rho = ir, \qquad \Theta = i\theta, \qquad \Phi = \phi,
\qquad \beta = -ib, \qquad \gamma = ic,
\end{equation}
which yields
\begin{equation}
ds^2 = -V(r)dt^2 + V(r)^{-1}dr^2 + r^2(d\theta^2 + \sinh^2\theta d\phi^2),
\label{hyperb}
\end{equation}
where $V(r)$ now is given by
\begin{equation}
V(r) = -1 - \frac{b(2-3bc)}{r} + 3bc + c r -k r^2.
\label{lapsehyp}
\end{equation}
For suitable values of the parameters, (\ref{hyperb}) describes a black
hole spacetime. The metric induced on the spacelike surfaces of constant
$r$ and $t$, in particular on the event horizon, is
\begin{equation}
d\sigma^2 = r^2(d\theta^2 + \sinh^2\theta d\phi^2).
\end{equation}
This is the standard metric of hyperbolic two-space $\mbox{H}^2$,
which has constant negative curvature and, of course,
is not compact. We now compactify the
$(\theta,\phi)$-sector by acting with an appropriate discrete
subgroup $G$ of the isometry
group $\mbox{SO(2,1)}$ of $\mbox{H}^2$, i.~e.~we consider the quotient
space $\mbox{H}^2/G$, which is compact. If we require it to be orientable,
it becomes a Riemann surface $S_g$ of genus $g>1$, and the topology
of the four-dimensional manifold is
${\hbox{{\rm I}\kern-.2em\hbox{\rm R}}}^2\times S_g$.
(For an exhaustive description of the compactification procedure
see e.~g.~\cite{balazs}).\\
Setting $c=0$ and $k=\Lambda/3 < 0$, $\Lambda$ being the cosmological
constant, one recovers the well-studied uncharged, static topological
black hole solutions in AdS gravity. Note however that now we have further
possibilities to construct black holes with non-trivial topology. E.~g.~we
may set $k=0$, yielding a solution which is not asymptotically AdS.
For $c>0$ and $-1 \le 3bc < 2$ it has a black hole interpretation,
$3bc = -1$ yielding an extreme black hole. The scalar curvature for $k=0$
is given by
\begin{equation}
R = -\frac{6c}{r}\left(1+\frac{b}{r}\right),
\end{equation}
hence we have a curvature singularity at
$r=0$. $R$ vanishes for $r \to \infty$. Note however that the
manifold is not Ricci-flat at infinity, therefore it does not approach
flat Minkowski space, and we have no contradiction to Hawking's theorem
\cite{hawk1}, because this makes essential use of asymptotic flatness.
In the case $k=0$, the
manifold is globally hyperbolic, in contrast to AdS black holes.\\
One further observes that (\ref{lapsehyp}) is equivalent to (\ref{lapse}),
which can easily be seen by setting
\begin{equation}
c = \gamma, \qquad b = -\beta + \frac{2}{3\gamma}.
\end{equation}
Consequently the lapse functions for the spherical and the $g>1$ black hole
are identical, a feature different from black holes in AdS gravity.
(Note however, that the full metrics (\ref{spher}) and (\ref{hyperb})
clearly are not equivalent).\\
Now set e.~g.~$b = 2/3c - \eta$ and $k=\Lambda/3>0$ in (\ref{lapsehyp}), giving
in the limit $c \to 0$
\begin{equation}
V(r) = 1 - \frac{2\eta}{r} - \frac{\Lambda r^2}{3}. \label{lapsedS}
\end{equation}
This represents a black hole spacetime similar to the Schwarz\-schild-de
Sitter solution, but now with non-trivial topology. The scalar curvature
approaches $4\Lambda$, as $r$ goes to infinity.
It is important to stress that (\ref{lapsedS}) is not a solution of
Einstein's equations $R_{\mu\nu} = \Lambda g_{\mu\nu}$, as in this case
a lapse function
\begin{equation}
V(r) = C - \frac{2\beta}{r} - \frac{\Lambda r^2}{3}
\end{equation}
would only admit horizons whose curvatures have the same sign as the constant
$C$, i.~e.~spherical ones for (\ref{lapsedS}).\\
For $9\beta^2\Lambda<1$ (\ref{lapsedS}) has two zeroes $r_-<r_+$,
$r_-$ representing the black hole's event horizon, whereas $r_+$ is
a cosmological horizon. The occurence of topological black holes
which are neither asymptotically flat nor AdS is also known in
dilaton gravity \cite{cai}.\\
In order to obtain a toroidal black hole spacetime, one can perform another
analytical continuation of (\ref{spher}) by
\begin{equation}
\tau = t\sqrt{d}, \qquad \rho = \frac{r}{\sqrt{d}},
\qquad \Theta = \theta\sqrt{d}, \qquad \Phi = \phi,
\qquad \beta = \frac{b}{\sqrt{d}}, \qquad \gamma = \frac{c}{\sqrt{d}}.
\end{equation}
Taking the limit $d \to 0$, we get
\begin{equation}
ds^2 = -V(r)dt^2 + V(r)^{-1}dr^2 + r^2(d\theta^2 + \theta^2 d\phi^2),
\label{tor}
\end{equation}
with
\begin{equation}
V(r) = \frac{3b^2c}{r} - 3bc + c r -k r^2.
\label{lapsetor}
\end{equation}
For suitable parameter values, (\ref{tor}) represents a black hole.
The $(\theta,\phi)$-sector carries now the flat metric
\begin{equation}
d\sigma^2 = r^2(d\theta^2 + \theta^2 d\phi^2).
\end{equation}
Passing from polar to cartesian coordinates according to
\begin{eqnarray}
x &=& \theta\cos\phi, \nonumber \\
y &=& \theta\sin\phi,
\end{eqnarray}
and identifying
\begin{eqnarray}
x &\simeq & x + n, \nonumber \\
y &\simeq & y + m,
\end{eqnarray}
with $n,m \in Z$, we get a compact orientable surface with genus $g=1$,
i.~e.~a torus, and the topology of our four-dimensional manifold
becomes ${\hbox{{\rm I}\kern-.2em\hbox{\rm R}}}^2\times S^1\times S^1$.\\
Setting $c=-2\eta/L$, $b=\sqrt{L/3}$, $k=\Lambda/3<0$, yields
in the limit $L \to \infty$
\begin{equation}
V(r) = -\frac{2\eta}{r} - \frac{\Lambda r^2}{3},
\end{equation}
which desrcibes for $\eta>0$ the uncharged static toroidal black hole known
from AdS gravity. Note that, unlike the $g>1$ case,
now for $k \ge 0$ one looses the black hole interpretation, as
(\ref{lapsetor}) does not have any real root for $k=0$, and only one real
root for $k > 0$, which is not a black hole event horizon.\\
Now one may doubt if the analytical continuations desribed above
yield again solutions of (\ref{fieldeq}); especially the limit
procedure to obtain the toroidal case may appear somewhat doubtful.
Therefore, following the derivation of the spherically symmetric case
in \cite{mannhkaz}, we make the ansatz
\begin{equation}
ds^2 = -f(\rho)dt^2 + g(\rho)d\rho^2 + \rho^2 d^2\Omega
\end{equation}
for the static topological black hole spacetime, where $d^2\Omega$
is given by
\begin{equation}
d^2\Omega = \left\{ \begin{array}{r@{\quad,\quad}l}
            d\theta^2 + \theta^2d\phi^2 & g = 1 \\
            d\theta^2 + \sinh^2\theta d\phi^2 & g > 1. \end{array} \right.
\end{equation}
An analogous calculation to \cite{mannhkaz} then yields exactly the
solutions above, namely (\ref{lapsehyp}) resp.~(\ref{lapsetor}),
confirming the validity of our analytical continuations.\\

\section{Thermodynamic Properties} \label{thermo}
Let us now discuss the thermodynamic properties of the
black hole solutions found above. First we remark that in the derivation
of the Hawking effect \cite{hawk2} the field equations do not enter,
so a black hole will emit Hawking radiation at temperature $\kappa/2\pi$
in an arbitrary theory of gravitation ($\kappa$ being the surface gravity).
Therefore it makes sense to assign the temperature $T=\kappa/2\pi$
with $\kappa = V'(r_H)/2$ to our topological black holes found above.
($r_H$ denotes the event horizon radial coordinate). As for the entropy,
Wald showed \cite{wald,waldiyer} that in an arbitrary theory of gravity
with diffeomorphism-invariant Lagrangian, one can define black hole
entropy as the Noether charge associated with the local symmetry
generated by the horizon Killing field. He further demonstrated the
validity of a first law of thermodynamics under the assumption of
asymptotic flatness. (See also \cite{jkm1,jkm2,koga} for black hole
thermodynamics in a generalized theory of gravity). However the topological
black holes in Weyl conformal gravity are not asymptotically flat,
so the validity of a first law remains to be verified. The definition
of the mass given in \cite{wald} is not applicable in the case
of non asymptotic flatness; a suitable
background subtraction has to be carried out. Yet, as we suspect the
entropy to be a local quantity closely connected to horizon properties,
it should be independent of the asymptotic behaviour. So we are led
to apply the Noether charge definition given in \cite{wald} also to
the topological black holes found above. As is shown in \cite{jkm1},
the entropy $S$ is then given by
\begin{equation}
S = - 2\pi\oint_{\Sigma} Y^{\mu\nu\rho\lambda}\hat{\epsilon}_{\mu\nu}
    \hat{\epsilon}_{\rho\lambda}\bar{\epsilon}, \label{entr}
\end{equation}
where
\begin{equation}
Y^{\mu\nu\rho\lambda} = \frac{\partial \tilde{L}}{\partial
                        R_{\mu\nu\rho\lambda}},
\end{equation}
$\tilde{L}$ denoting the scalar Lagrangian density ($I=\int d^4x \sqrt{-g}
\tilde{L}$). $\hat{\epsilon}_{\mu\nu}$ is the binormal to the (arbitrary)
cross-section $\Sigma$ of the horizon, and $\bar{\epsilon}$ the induced
volume form on $\Sigma$. For the action (\ref{act}), $Y^{\mu\nu\rho\lambda}$
reads
\begin{eqnarray}
Y^{\mu\nu\rho\lambda} &=& \alpha(g^{\mu\rho}R^{\nu\lambda} - g^{\nu\rho}
                          R^{\mu\lambda} - g^{\mu\lambda}R^{\nu\rho} +
                          g^{\nu\lambda}R^{\mu\rho}) \nonumber \\
                     && - \frac{2\alpha}{3}R^{\beta}_{\; \beta}
                          (g^{\mu\rho}g^{\nu\lambda} -
                          g^{\nu\rho}g^{\mu\lambda}).
\end{eqnarray}
Now consider e.~g.~the $g>1$ black hole (\ref{hyperb})
with lapse function (\ref{lapsedS}) and $\Lambda = 0$. Using the
theorem of Gauss-Bonnet, the entropy (\ref{entr})
is then easily calculated, and reads
\begin{equation}
S = \frac{256\pi^2\alpha}{3}(g-1).
\end{equation}
Hence we obtained a result which is independent of the parameter
$\beta$, and therefore the variation $\delta S$ vanishes identically.
The situation becomes even more drastically if one considers the
Kerr solution. Being Ricci-flat, it still satisfies the field equations
of Weyl conformal gravity. Furthermore it is asymptotically flat, so
the definitions of mass and angular momentum given in \cite{wald} are
applicable. As $R_{\mu\nu} = 0 = R^{\nu}_{\; \nu}$,
$Y^{\mu\nu\rho\lambda}$ is zero
in this case. The Noether current is proportional to $Y$ \cite{jkm1},
hence it also vanishes. This implies that the entropy, mass and
angular momentum of the Kerr black hole in Weyl conformal gravity
are all zero, which means that the first law is satisfied in a trivial
way. The vanishing of all these quantities certainly appears to be puzzling,
and is closely related to the absence of an Einstein-Hilbert term
and a fundamental length scale in the action (\ref{act}).
However, in \cite{hassl} it was shown how one can still obtain the
Bekenstein-Hawking entropy for the Schwarz\-schild black hole 
(at least at leading order) in
Weyl gravity. One considers the breaking at quantum level of
conformal invariance, generating thereby an effective Einstein-Hilbert
action (see also \cite{mott}). In this approach the gravitational
constant arises as
an induced quantity in the spirit of Sakharov \cite{sakh}; it is
caused by the quantum fluctuations of matter fields in the vacuum.
Let us shortly review how this works. The mentioned quantum fields $\Phi$
propagating in an external gravitational background $g$
induce an effective (euclidean) action $I_{ind}[g]$ according to
\begin{equation}
\exp(-I_{ind}[g]) = \int {\cal D}\Phi \exp(-I_m[\Phi,g]), \label{Iind}
\end{equation}
where $I_m[\Phi,g]$ is the action of the matter fields. The total partition
function $Z_{tot}$ reads
\begin{equation}
Z_{tot} = \int {\cal D}g \exp(-I_g[g]-I_{ind}[g]), \label{Ztot}
\end{equation}
$I_g[g]$ being the Weyl gravitational action given by (\ref{act}).
One can now show \cite{hassl,mott} (see also \cite{frol1,frol2,frol3})
that $I_{ind}[g]$ contains an Einstein-Hilbert term. In order to
obtain the black hole entropy, one calculates (\ref{Ztot}) in the
saddle point approximation. For the Schwarz\-schild black hole this yields  
the Bekenstein-Hawking entropy, resulting from the action (\ref{Iind}),
plus correction terms, coming from (\ref{act}). The Bekenstein-Hawking
entropy is pure entanglement in this approach, it is generated by matter
states hidden by an event horizon. (Interesting results on black hole
entropy in induced gravity can also be found in \cite{jac,frol1,frol2,frol3}).
Note however that the topological black hole solutions found in section
\ref{sol}
do not solve the field equations following from the Einstein-Hilbert
action. Therefore a calculation of the entropy as was done above
for the Schwarz\-schild case would not yield the Bekenstein-Hawking
result. Yet, if the quantum fluctuations induce also an effective
cosmological constant (it is shown in \cite{frol1} how this could work),
the class of Einstein spaces among the topological black holes considered
in the present paper are still solutions of the field equations following
from the induced action. So one would obtain an entropy which is
one quarter of the event horizon area, this part being pure entanglement,
plus correction terms.
In any case, the induced gravity approach to the
entropy of the new black hole solutions found above seems to be more
promising than the Noether charge method, which fails here due to the
absence of a length scale.

\section{Summary and Discussion}
\label{disc}
In this paper we derived exact solutions of Weyl conformal gravity
representing topological black holes. We found a generalization of
the topological black holes which occur in AdS gravity. For genus $g>1$ this
generalization includes spacetimes which are not asymptotically AdS,
are globally hyperbolic and have vanishing scalar curvature at infinity.
Moreover topological black holes with $g>1$ can also occur for a positive
cosmological constant. On the other hand, in the toroidal case, only
solutions which are asymptotically AdS occur.\\
As far as thermodynamical quantities are concerned, it makes sense to
assign a temperature $T = \kappa/2\pi$ to our black hole solutions,
because the field equations do not enter the derivation of the Hawking
effect. However,
the idea of black hole entropy as a Noether charge pioneered by Wald, and
applicable to an arbitrary theory of gravity with diffeomorphism-invariant
Lagrangian, does not yield satisfactory results in the case of Weyl
conformal gravity, because the first law seems to be satisfied in a
trivial way. This suggests that entropy and mass of black holes
in Weyl gravity should be treated in the sense of \cite{hassl},
i.~e.~one does not assign any entropy to the classical gravitational field;
rather it is generated by quantum matter states hidden by the event
horizon.\\
Last, it seems worth mentioning that
besides the fact that Weyl conformal gravity like AdS gravity admits black
holes with unusual topology, these two theories of gravitation possess
another common quality, being the only types of gravity which have a
consistent interaction with massless higher spin fields \cite{frad}.\\

\section*{Acknowledgement}

This work has been supported by a research grant within the
scope of the {\em Common Special Academic Program III} of the
Federal Republic of Germany and its Federal States, mediated 
by the DAAD.\\ 
The author would like to thank L.~Vanzo for helpful discussions.

\end{document}